\documentclass[12pt]{iopart}

\usepackage{graphicx,iopams,epsf}
\begin{document}
\epsfclipon

\title[Finite-size scaling and fluctuation effects in the Kuramoto model]
{To synchronize or not to synchronize, that is the question: finite-size scaling and fluctuation effects in the Kuramoto model}

\author{Lei-Han Tang}

\address{Department of Physics, Hong Kong Baptist University, Kowloon Tong, Hong Kong SAR, China}
\ead{lhtang@hkbu.edu.hk}
\begin{abstract}
The entrainment transition of coupled random frequency oscillators presents a long-standing 
problem in nonlinear physics. The onset of entrainment in populations of large but finite size exhibits 
strong sensitivity to fluctuations in the oscillator density at the synchronizing frequency. This is the 
source for the unusual values assumed by the correlation size exponent $\nu'$. Locally coupled 
oscillators on a $d$-dimensional lattice exhibit two types of frequency entrainment: 
symmetry-breaking at $d > 4$, 
and aggregation of compact synchronized domains in three and four dimensions. 
Various critical properties of the transition are well captured by finite-size scaling relations with
simple yet unconventional exponent values.
\end{abstract}

\maketitle

\section{Introduction}

Mathematical modeling of coherent oscillation in a large population of autonomous units 
has a long and interesting history, as discussed in several monographs\cite{ref:Winfree,ref:Kuramoto} and 
recent reviews\cite{ref:Strogatz,ref:Acebron,ref:Mendes}. 
One particular paradigm is the Kuramoto model of coupled random frequency oscillators\cite{ref:Kuramoto}. 
In terms of the phase variables $\phi_j, j=1,\ldots, N$, the dynamical equations take the form,
\begin{equation}
{d\phi_j\over dt}=\omega_j-K\Delta\sin(\phi_j-\theta).
\label{mf-phi}
\end{equation}
Here $K$ is the coupling strength among the oscillators, and $\Delta (>0)$ and $\theta$ are the
amplitude and phase of the mean instantaneous phase factor,
\begin{equation}
\Delta e^{i\theta}\equiv N^{-1}\sum_{j=1}^N e^{i\phi_j},
\label{eq:delta}
\end{equation}
which is also known as the order parameter for entrainment.
The intrinsic frequencies $\omega_j$ are drawn from a distribution $g(\omega)$. 
In this paper we shall focus on the case where $g(\omega)$ is a gaussian function
centered at $\omega_{\rm max}=0$ with unit variance.

In the classic work by Kuramoto\cite{ref:Kuramoto}, existence of an entrained state is
established by considering solutions to (\ref{mf-phi}) at a constant $\Delta$. 
Under this assumption, Eq. (\ref{mf-phi}) is easily integrated.
After an initial transient, oscillators with $|\omega_j|<K\Delta$ each
reaches a fixed angle and together form the entrained group, while those with 
$|\omega_j|>K\Delta$ go into periodic motion at modified frequencies $\tilde{\omega}_j=\sqrt{\omega_j^2-(K\Delta)^2}$.
In the limit $N\rightarrow\infty$, only the entrained population contributes to the sum (\ref{eq:delta}).
The resulting self-consistent condition
\begin{equation}
\Delta=\int_{-K\Delta}^{K\Delta}d\omega g(\omega)\sqrt{1-(\omega/K\Delta)^2}
\label{Delta-mf}
\end{equation}
has a nontrival solution $\Delta>0$ when $K>K_c=2/\pi g(0)$. 
Thus entrainment requires the product $Kg(0)$ to exceed a threshold value.
On the supercritical side, $\Delta(K)\sim (K-K_c)^\beta$ with the usual mean-field exponent
$\beta=1/2$.

While the above predictions agree well with numerical solutions of the globally coupled Kuramoto model, 
there are several unresolved issues regarding the stability of the mean-field solution, 
as discussed elegantly in Ref. \cite{ref:Strogatz}. 
From a physicist's point of view, one would like to understand how temporal fluctuations of $\Delta(t)$,
which are present when the sum (\ref{eq:delta}) contains only a finite number of terms,
affect the behavior predicted by the mean-field treatment.
When the oscillator frequencies $\omega_j$ are drawn independently from the distribution $g(\omega)$,
sample-to-sample fluctuations in the time averaged $\Delta$ need to be considered as well. 
These effects culminate at $K=K_c$ where the size of the entrained group grows sub-linearly
with the system size $N$.

Daido developed a perturbation theory\cite{ref:Daido90} to calculate the susceptibility defined by,
\begin{equation}
\chi\equiv N\bigl(\overline{\Delta^2}-\overline{\Delta}^2\bigr)
\label{eq:chi0}
\end{equation}
Here the overline bar denotes time average. His calculation yielded a power-law divergence of
$\chi$ at $K_c$, but the exponent $\gamma$ differs on the two sides of the transition:
$\gamma=\gamma_-=1$ on the subcritical side and $\gamma=\gamma_+=1/4$ on the supercritical side.
Daido has further suggested a finite-size scaling form based on his own numerical studies,
\begin{equation}
\chi=|k|^{-\gamma_\pm}\Phi(kN^{1/\nu_\pm'}).
\label{eq:chi}
\end{equation}
Here $k=(K-K_c)/K_c$ is the distance to the transition, $\nu_-'=2$ and $\nu_+'=1/2$. These 
exponent values unfortunately do not agree with later numerical work\cite{ref:Hong02}.

Equations (\ref{mf-phi}) have been adapted to describe synchronization phenomena in finite dimensions with local 
oscillator coupling\cite{ref:Sakaguchi},
\begin{equation}
{d\phi_j\over dt}=\omega_j-K\sum_{l\in\Lambda_j}\sin(\phi_j-\phi_l).
\label{d_dim-phi}
\end{equation}
Here $\Lambda_j$ denotes the set of the nearest neighbor sites of $j$ on a $d$-dimensional hypercubic lattice.
Arguments have been presented to show that, as the linear system size $L\rightarrow\infty$, global
entrainment is not possible when $d<2$\cite{ref:Sakaguchi}. 
Analytical and numerical studies of the model in three to six dimensions by Hong, Chat\`e, Park and Tang (HCPT)\cite{ref:HCPT} 
have established the following.
For $d>4$, the entrainment transition is of the symmetry-breaking type
with exponents indicative of a mean-field behavior.
At $d=3$ and 4, on the other hand, global entrainment is reached through aggregation of locally synchronized domains
as the coupling strength $K$ increases beyond a critical value $K_c$. Although the aggregation process is continuous
and can be associated with a length scale $\xi$ that diverges at $K_c$, the fraction of entrained oscillators may
undergo a discontinuous jump at $K_c$ in the limit $L\rightarrow\infty$. 

Below we shall give an overview of our current knowledge on the finite-size scaling properties 
in the all-to-all coupled Kuramoto model and discuss
some fine differences between this case and the Kuramoto model on a random graph. We shall argue that the latter 
provides the correct mean-field description of the entrainment transition at $d>4$. Key features of the entrainment
process in three and four dimensions are also discussed. However, no attempt is made to review the vast amount
of related work in the literature. Fortunately, this role is partially fulfilled by several recent articles cited above.

\section{Finite-size scaling in the globally coupled model}

\subsection{Anomalous behavior due to quenched frequency fluctuations}

In typical numerical studies of (\ref{mf-phi}), the frequencies $\omega_j$ are drawn independently from the
distribution $g(\omega)$. HCPT have shown that, in this case, $\nu'=5/2$ on either side of the transition.
Since the argument is quite straightforward and relevant for the discussion of synchronization
on complex networks and in finite dimensions, we reproduce it below.

Still assuming $\Delta$ in (\ref{mf-phi}) to be time-independent and restricting the sum in (\ref{eq:delta})
to the entrained oscillators in a finite population, Eq. (\ref{Delta-mf}) is replaced by
\begin{equation}
\Delta=\Psi(\Delta) \equiv {1\over N}\sum_{j,|\omega_j|<K\Delta}\sqrt{1-\Bigl({\omega_j\over K\Delta}\Bigr)^2}.
\label{Delta_tilde}
\end{equation}
Since the actual values of the $\omega_j$'s vary from sample to sample,
the sum $\Psi$ has a ``quenched'' fluctuation 
$\delta\Psi\equiv\Psi-\langle\Psi\rangle
\propto N_s^{1/2}/N$, where
$N_s$ is the number of oscillators in the frequency interval 
$(-K\Delta,K\Delta)$. Here $\langle\cdot\rangle$ denotes average over different realizations of the random frequencies.
Close to the transition, an expansion of Eq. (\ref{Delta_tilde}) at small $\Delta$ yields,
\begin{equation}
\Delta = (K/K_c)\Delta - c(K\Delta)^3+\delta\Psi,
\label{delta_N}
\end{equation}
where $c=-\pi g''(0)/16$. Mathematically,
the variance of $\delta\Psi$ can be computed from (\ref{Delta_tilde}) for independently drawn frequencies:
$\langle(\delta\Psi)^2\rangle=4g(0)K\Delta/3N+O(\Delta^2/N)$.
Thus solution to (\ref{delta_N}) takes the scaling form,
\begin{equation}
\Delta(K,N)=N^{-1/5}f(kN^{2/5}).
\label{delta-scaling}
\end{equation}
The scaling function $f(x)$ is sample-dependent and satisfies the equation,
\begin{equation}
xf-cK_c^3f^3+(8/3\pi)^{1/2}\epsilon f^{1/2}=0.
\label{f-eqn}
\end{equation}
Here $\epsilon\equiv\delta\Psi/\langle(\delta\Psi)^2\rangle^{1/2}$
is a gaussian random variable with zero mean and unit variance.

Equation (\ref{delta-scaling}) suggests $\nu'=5/2$ as seen in numerical experiments.
Its origin lies in the fluctuations in the oscillator density at the peak frequency $\omega_{\rm max}$ of $g(\omega)$.
This effect produces a sample-dependent shift $\delta K_c\sim N^{-2/5}$ of the entrainment threshold. 
At the nominal threshold $K=K_c$, $\Delta_0\sim\delta K_c^\beta\sim N^{-1/5}$. In other words, the
number of entrained oscillators right at $K=K_c$ scales as $N^{4/5}$.
(Note that, when the term $\epsilon$ in (\ref{f-eqn}) is negative, there is only the trivial
solution $f=0$ at $x=0$. For these samples, the entrained population size is much smaller than $N^{4/5}$.)
Numerical simulations of (\ref{f-eqn}) produced values of $\langle\Delta^2\rangle$ in quantitative agreement with that
of the Kuramoto model for $K>K_c$\cite{ref:HCPT}.

\subsection{Dynamic fluctuations in the subcritical region}

For $K<K_c$, oscillators essentially run with their intrinsic frequencies. If the phase factors in the 
sum (\ref{eq:delta}) were all statistically independent from each other, one would have $\overline\Delta^2=1/N$ 
and hence $\chi=1$.
The perturbative calculation by Daido\cite{ref:Daido90} yielded the following expression for $K$ close to $K_c$:
\begin{equation}
\chi={A\over K_c-K}+D.
\label{curie}
\end{equation}
Here $A=4/K_c$ for the gaussian $g(\omega)$, and
$D$ is a numerical constant which can also be evaluated, though the calculation is rather tedious.
Equation (\ref{curie}) was rederived recent by Hildebrand {\it et al.} using a different approach\cite{ref:Chow}. 

Numerical results by Hong {\it et al.}\cite{Hong_unpub} are in good agreement with (\ref{curie})
when the system size $N>N_c\sim |K-K_c|^{-5/2}$. At smaller values of $N$, a sample-dependent behavior is seen,
so the result depends on the averaging procedure. For example, the data for $\chi$ calculated from 
(\ref{eq:chi0}) for a given sample and then averaged over many samples obey the scaling 
$\langle\chi\rangle\simeq |k|^{-1}\Psi(kN^{2/5})$. This implies that, in the critical region $|k|<N^{-2/5}$,
dynamic fluctuations of $\Delta$ have an amplitude $\delta\Delta_{\rm dyn}\sim N^{-3/10}$, which is weaker than
the sample to sample fluctuation $\Delta_0\sim N^{-1/5}$. Due to this difference,
the hyperscaling relation $\nu'=2\beta+\gamma$ is violated. At present, there is no analytic theory that explains
these numerical findings.

\subsection{Finite-size scaling in the ``pure'' case}

The oscillator density fluctuations on the frequency axis responsible for the anomalous $\nu'=5/2$ can be eliminated 
when the frequency $\omega_j$ of the $j$'th oscillator is chosen according to the formula
\begin{equation}
{j-0.5\over N}=\int_{-\infty}^{\omega_j}g(\omega)d\omega.
\label{frequency-select}
\end{equation}
The intrinsic frequencies of the oscillators in this case are nearly uniformly spaced locally on the axis.
In analogy with the discussion of disordered systems, we call this situation the ``pure'' case.

Numerical simulations of the Kuramoto model with frequencies given by (\ref{frequency-select})
produced exactly the same $\Delta(K)$ function in the large size limit as the ``disordered'' case.
On the other hand, measurement of fluctuations of $\Delta(t)$ yielded
$\gamma_\pm=1/4$ and $\nu_\pm'=5/4$ which are very different from the
disordered case\cite{Hong_unpub}. It thus appears that the subcritical behavior (\ref{curie}) requires extra assumptions
which are not fulfilled in the pure case.

\subsection{Entrainment transition on complex networks}

The entrainment transition of the Kuramoto model has been studied on complex networks by several groups\cite{ref:Mendes}.
Here the dynamical equations are given by (\ref{d_dim-phi}), with $\Lambda_j$ denoting the set of nodes
connected to node $j$ on the network. In the case of scale-free networks where the degree $k$ of nodes satisfies
a power-law distribution $P(k)\sim k^{-\alpha}$, both numerical and analytical studies show that 
the exponent $\beta=1/2$ for $\alpha\geq 5$, but increases sharply as $\beta=1/(\alpha-3)$ for 
$3<\alpha<5$\cite{ref:Lee}. The size of the critical region where finite-size effects become significant, as described by
the exponent $\nu'$, also broadens in the latter case. These two effects combine to give a broadened
transition when the local connectivity of the network assumes a broad distribution.

Interestingly, for $\alpha>5$, and on the random graph with a fixed degree, 
the finite-size scaling form (\ref{eq:chi}) is found to hold across the transition, with the
exponents given by $\gamma_\pm=3/2$ and $\nu_\pm'=5/2$\cite{ref:Hong07}. This is in contrast to the behavior of the
Kuramoto model with all-to-all coupling. To better illustrate the difference, let us examine
the distribution of the mean phase velocity $v_j=\overline{d\phi_j/dt}$ of oscillators in the neighborhood
of the transition. Figure 1(a) shows a comparison of the cumulative distributions obtained from 
the all-to-all coupling case and on a random graph at $k=6$. 
Both systems are in the critical region of their respective entrainment transitions. 
Entrained oscillators have the same phase velocity $v_{\rm syn}\simeq 0$ as indicated by the vertical jumps of the curves. 
Distinct behavior is seen right above and below $v_{\rm syn}$ in the two cases.
For all-to-all coupling, the curve is nearly flat, suggesting a vanishing density of oscillators in the neighborhood of
$v_{\rm syn}$. Thus the entrained population is well-separated from the running oscillators.
On the random graph, the vertical jump joins smoothly with the
detrained regions. The rise at small $|v|$ can be fitted to a $|v|^{1/2}$-law. Consequently, the 
oscillator density near $v_{\rm syn}$ diverges as $|v|^{-1/2}$. The boundary between the
entrained and running populations is much less well-defined in this case.

The difference in the density of oscillators near the entrainment frequency can be appreciated from
the following consideration. For all-to-all coupling, the fluctuation effect on a given oscillator is due to 
temporal variations in the system-averaged order parameter $\Delta$ which vanishes in the large size limit.
In addition, from the scaling $\Delta_{\rm dyn}\sim N^{-3/10}$ and $\Delta_0\sim N^{-1/5}$
mentioned above, $\Delta$ can be regarded as a constant in the first approximation even in the critical
region. The mean phase velocity of oscillators then essentially follows the behavior predicted by
Kuramoto's mean-field analysis, which is what the black curve in Fig. 1(a) suggests. In contrast,
on the random graph, each oscillator is coupled to a small number of other oscillators.
Hence the fluctuation effect has a finite strength and does not decrease significantly in the infinite size limit.
Oscillators with the ``right environment'' first become entrained, but due to the presence of fluctuations,
they undergo diffusive phase motion in a weak ordering field $\Delta$, punctuated with phase slips.
This diffusive background is present even on the entrained side.

\begin{figure}
\epsfxsize=\linewidth
\epsfbox{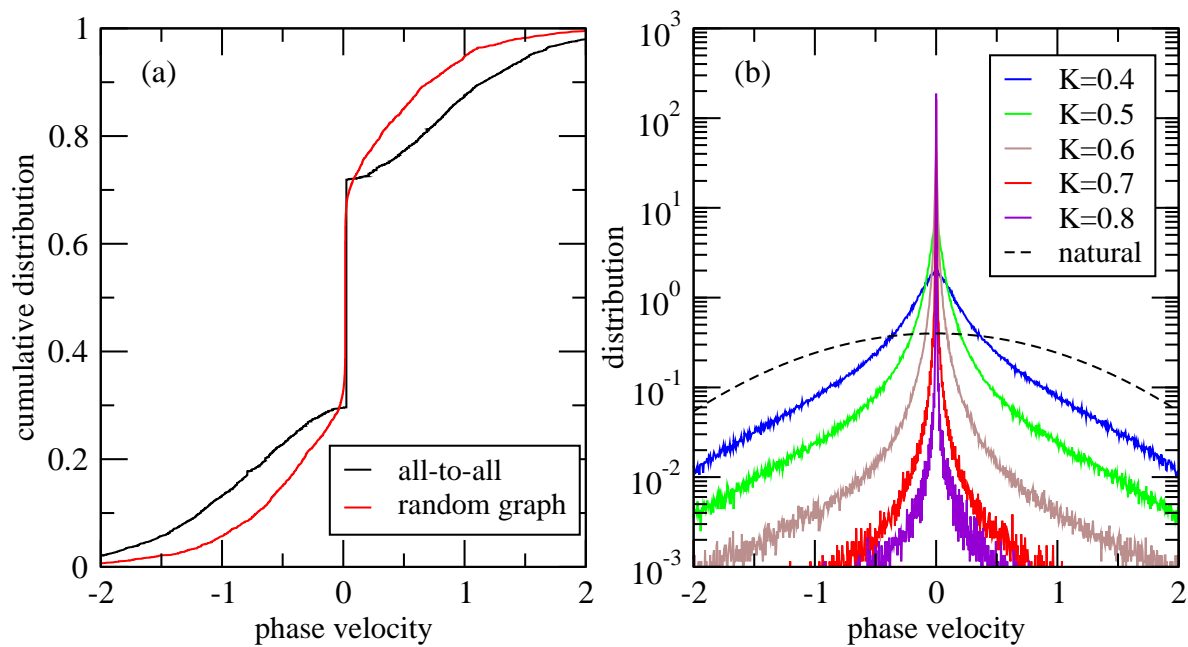}
\caption{(Color online) (a) The cumulative distribution of the phase velocity at criticality for 
all-to-all (black, $K=\sqrt{8/\pi}$) and random-graph (red, $K=0.38$, degree = 6) couplings. Here $N=3200$ in both cases.
(b) The phase velocity distribution in a given sample in three dimensions at different coupling strength $K$.
Here $N=128^3$ and $K_c\simeq 0.7$. Dashed line indicates the frequency distribution at $K=0$.
}
\label{fig:freq_dis}
\end{figure}

\section{The Kuramoto model in finite dimensions}

The ``quenched'' random frequencies in (\ref{d_dim-phi}) introduce a static phase deformation $\phi_j^{(0)}$
whose spatial structure has been argued to be responsible for the different types of entrainment 
behavior in finite dimensions\cite{ref:HCPT}.
Salient features of this static phase field are captured by a linear analysis carried out by
Sakaguchi {\it et al.}\cite{ref:Sakaguchi} and by Hong {\it et al.}\cite{ref:Hong05}  The resulting dynamic equations,
which reduces to the ``quenched Edwards-Wilkinson equation''\cite{ref:Edwards} in the continuum limit,
exhibit a diffusive relaxation towards a ``fully entrained'' state with vanishing phase velocity for all oscillators. 
As the linear system size $L\rightarrow\infty$, fluctuations of $\phi_j^{(0)}$ remain bounded when the
system dimension $d>4$, but diverge when $d\leq 4$. In addition, the gradient of $\phi_j^{(0)}$ diverges
when $d\leq 2$.

HCPT proposed a plausible scenario for the entrainment transition based on the linear analysis.
The linear approximation is justified at sufficiently large $K$ and above the lower critical dimension $d_l=2$,
where the run-away oscillators are absent. (We assume here that the intrinsic frequencies are bounded.) 
As the coupling strength $K$ decreases,
more and more oscillators break away from entrainment. The question is whether the remaining oscillators
can stay phase-locked as their fraction in the population diminishes. For $d>4$, the entrained state
has bounded phase fluctuations and hence breaks the phase symmetry $\phi\rightarrow\phi+c$.
In this case, $\Delta$ as defined by Eq. (\ref{eq:delta}) can still be used as a global order parameter 
and acts as a spontaneous ordering field on the oscillators as in the globally coupled case.
On the other hand, for $d\leq 4$, the phase symmetry is not
broken. Therefore phase-locking need to be maintained via a percolative network of entrained oscillators
in space. The latter requires a finite density of these oscillators given the random assignment of the
oscillator frequencies. The entrainment transition, when viewed in terms of the fraction of entrained
oscillators in the system, is expected to be discontinuous.

The above picture is largely confirmed by extensive simulations carried out by HCPT in three to six dimensions
which also yielded additional details of the entrainment transition\cite{ref:HCPT}. In five and six dimensions, the
distribution of the mean phase velocity shows essentially the same behavior as the Kuramoto model on
a random graph as depicted in Fig. 1(a). On the other hand, a very different distribution of phase velocities is
obtained in three and four dimensions. Figure 1(b) presents an example
from a $N=128^3$ system in three dimensions. Significant narrowing of the distribution takes place 
at $K$ values well below the threshold $K_c\simeq 0.7$ at this size. 
The wings of the distribution can be fitted to a power-law $P(v)\sim |v|^{-2}$.
The data is consistent with the picture that, below $K_c$,
entrainment takes place locally through the formation of entrained clusters. 
The size of the clusters grow as $K$ increases, reaching the system size at  $K_c$. 
The increase of the entrainment threshold with the system size, as observed in simulations,
can be taken as further confirmation of this picture\cite{ref:HCPT}.

HCPT also examined various singular behavior of the system at the entrainment transition with the help of finite-size scaling
analysis. For $d>4$, $\Delta$ as defined by Eq. (\ref{eq:delta}) is
a suitable order parameter. For $d=3$ and 4, static phase fluctuations  in
a sufficiently large system render $\Delta$ vanish even on the entrained side. 
Instead, the Edwards-Anderson order parameter
\begin{equation}
\Delta_{\rm EA}=\lim_{t-t_0\rightarrow\infty}{1\over N}\Bigl|\sum_je^{i[\phi_j(t)-\phi_j(t_0)]}\Bigr|
\label{delta_ea}
\end{equation}
was found to be appropriate. Numerical results for a range of system sizes
are in good agreement with the finite-size scaling
\begin{equation}
\langle\Delta^2\rangle, \langle\Delta^2_{\rm EA}\rangle =L^{-2\beta/\nu}\tilde\Phi(kL^{1/\nu}).
\end{equation}
The exponent values extracted from the simulation data can be expressed as,
\begin{eqnarray}
d&\leq&4:\qquad \beta=0,\quad \nu=2/(d-2);\nonumber\\
d&>&4:\qquad \beta=1/2,\quad \nu=5/(2d),\quad \nu'\equiv d\nu=5/2.
\label{exponents}
\end{eqnarray}
Note that in three and four dimensions, $\beta=0$ is consistent with a jump in 
the fraction of entrained oscillators at the transition. 

\section{Conclusions and outlook}

Numerical investigations of the Kuramoto model with all-to-all coupling, on scale-free networks and random graphs, and on
finite dimensional lattices, revealed a wealth of behavior at the onset of global entrainment. The usual finite-size
scaling approach has been shown to be successful in capturing the key critical properties of these systems, 
including fluctuation effects and the size of the entrained population at the transition.
Development of phenomenological arguments based on various approximations enhanced our understanding
of the entrainment process. However, a full analytic theory of fluctuations is still lacking,
even in the case of the Kuramoto model with all-to-all coupling.

The original Kuramoto model (\ref{mf-phi}) has a small parameter $\Delta$ in the neighborhood of the transition
which can be employed for systematic expansions. However, such a treatment need to be handled
with care when considering oscillators with $\omega_j$'s comparable to $\Delta$. These oscillators are in fact the
ones that become entrained first and contribute significantly to $\Delta$. The good news is that, as Fig. 1(a) shows,
temporal fluctuations of $\Delta$ are weak when compared to $\Delta$ itself at sufficiently large $N$.
Therefore one may try to adapt Daido's scheme\cite{ref:Daido90} on the supercritical side to a finite system at $K=K_c$ to 
compute the size of the entrained population, from which the exponent $\nu'$ can be obtained.
This work is in progress.

For the Kuramoto model on random graph and on finite-dimensional lattices with $d>4$, the above perturbative
scheme can not be applied directly. However, since the entrainment transition is of the symmetry breaking type,
a coarse-grained description is expected to be appropriate. For example, one may consider equations of the complex
Ginzburg-Landau type with suitable thermal and quenched disorder terms that mimic the effects of the random
frequencies. Standard field theoretic techniques can then be applied to perform the relevant calculations.

Existing analytic methods may not be able to treat the entrainment transition in three and four dimensions.
To describe in detail the breakdown of entrained clusters as the coupling strength $K$ weakens,
one needs to know the dynamic behavior of defects (grain boundaries and dislocations) that mediate phase slips
between neighboring entrained domains. Establishment of relevant scaling relations will help one to sort out various types of
complex spatiotemporal correlations, from which a better understanding of the entrainment process can be expected.

{\bf Acknowledgement:} Many of the ideas presented here grew out of an enjoyable collaboration with Hugues Chat\`e,
Hyunsuk Hong, and Hyunggyu Park. 
The work is supported in part by the Research Grants Council of the HKSAR under grant 202107.

\end{document}